\documentclass[]{spie}  

\usepackage{graphicx}
\graphicspath{{Figures/}}
\DeclareGraphicsExtensions{.eps,.ps}

\usepackage{amsmath}
\usepackage{amsfonts}
\usepackage{amssymb}

\title{Fluctuational Instabilities of Alkali and Noble Metal Nanowires}

\author{J.~B\"urki, C.A.~Stafford, and  D.L.~Stein
\skiplinehalf
Department of Physics, University of Arizona, 1118 E. 4th St., Tucson, AZ, USA
}

\begin{document}
  \maketitle

\begin{abstract}
We introduce a continuum approach to studying the lifetimes of monovalent metal
nanowires.  By modelling the thermal fluctuations of cylindrical
nanowires through the use of stochastic Ginzburg-Landau classical field
theories, we construct a self-consistent approach to the
fluctuation-induced ``necking'' of nanowires.  Our theory provides
quantitative estimates of the lifetimes for alkali metal nanowires
in the conductance range $10 < G/G_0 < 100$ (where $G_0=2e^2/h$ is the
conductance quantum), and allows us to account for qualitative differences in the
conductance histograms of alkali vs.\ noble metal nanowires.
\end{abstract}


\section{INTRODUCTION}
\label{sec:intro}

A metallic nanowire, defined as a wire whose thinnest cross-section
contains only a few, to a few hundred, atoms, acts essentially as an
incompressible fluid of electrons, at least for simple monovalent metals,
such as the alkali or noble metals.  One would therefore expect a long
cylindrical nanowire to break apart due to the Rayleigh instability. Long
gold nanowires have nevertheless been observed in transmission electron
microscopy (TEM) experiments\cite{kondo97,kondo00}, and in fact appear to
be surprisingly stable, with lifetimes of the order of seconds.  This
apparent paradox has been resolved theoretically\cite{kassubek00,zhang02a,urban03} with the inclusion of quantum-size effects, which have been shown to stabilize the wires
for a set of ``magic'' radii. The stability thus arises through a
competition of shell-effects, comparable to what happens in metal clusters,
and an interplay of Rayleigh and Peierls instabilities.  Linear stability
analyses have shown cylindrical nanowires to be very stable, while a
dynamical model, including surface self-diffusion, indicates that they
should form spontaneously from random initial wires\cite{burki03}.

Despite their robustness, cylindrical wires are nevertheless only
metastable, with finite lifetimes in the presence of thermal fluctuations.
In this paper, we introduce a continuum approach to studying the lifetimes
of monovalent metal nanowires. The ionic medium is treated as an
incompressible continuum (jellium), and electron-shell effects are
treated semiclassically.  This approach appears very promising for studying
nanowires of ``intermediate'' thickness, which are thin enough that
electron-shell effects play a dominant role, but not so thin that a
continuum approach is unjustified.
Thermal fluctuations are modelled through
the use of stochastic Ginzburg-Landau classical field theories, and we
construct a self-consistent approach to the fluctuation-induced
``necking'' of nanowires.
Our theory provides quantitative estimates
of the lifetimes for alkali nanowires with electrical conductance $G$ in
the range $10<G/G_0<100$, where $G_0=2e^2/h$ is the conductance quantum.
Moreover, our theory accounts qualitatively
for the large difference in the observed
stability of alkali vs.\ noble metal nanowires.

The paper is organized as follows: In section~\ref{sec:nanowires}, we give
a brief survey of the phenomenology of nanowires, with particular emphasis
on the results that are relevant to the lifetime of nanowires. The
continuum free-electron model is presented in section~\ref{sec:model},
while sections~\ref{sec:escape}-\ref{sec:prefactor} describe the field theory
from which the lifetime is computed. Finally, section~\ref{sec:discussion} is devoted to
discussion and analysis of the results, with comparison to existing
experiments.

\section{PHENOMENOLOGY OF NANOWIRES}
\label{sec:nanowires}

Metallic nanowires have been extensively studied in the past
decade\cite{Agrait93,rubio96,stalder96,Brandbyge95,Krans95,costa-kramer97b,yanson98,Agrait03}.
Originally, short thin contacts were formed by crashing the gold tip of a
scanning tunneling microscope (STM) into a gold sample. Upon subsequent
retraction of the tip, the contact gradually necked down until it broke,
and the electrical conductance through the contact was recorded. An atomic
force microscope was used to simultaneously measure the tensile force
applied to the contact. Conductance plateaus at integer multiples of
$G_0$ were observed, together with a sawtooth behavior of the force,
with a perfect correlation between abrupt changes in both quantities.
Because of the inherent irreproducibility of the measurements -- wires have
a different structure on each cycle -- and the imperfection of the
conductance quantization, statistical analyses have turned out to be very
useful. A different setup, the mechanically controllable break junction
(MCBJ)\cite{muller92}, has been used in most statistical studies. In this
technique, a notched wire is glued to a substrate, whose bending is used to
deform the contact. By breaking the wire and putting it back together, one
can form a nanocontact, which can be repeatedly broken and reformed.
Conductance histograms\cite{Brandbyge95,Krans95,costa-kramer97b,yanson98},
built out of thousands of conductance traces, have clear peaks at positions
close to, but below, integer multiples of the conductance quantum $G_0$.
Which peaks are present depends on the metal considered: For example, gold
has all peaks $1, 2, 3, 4, \dots G_0$\cite{Brandbyge95,costa-kramer97b},
while sodium has large peaks at $G=1,3,5,6,\dots G_0$\cite{Krans95},
with only smaller peaks at $G=2,4 \dots G_0$.\cite{urban04b}
The shift and
broadening of the peaks have been shown to be due to
disorder\cite{garcia-mochales97,burki99,burki01} both in the contact and
the leads, either in the form of impurities, or more likely, of surface
corrugation due to the imperfect atomic structure.

More recently, the MCBJ technique has been used to build histograms for
alkali metal nanocontacts with larger
conductances\cite{yanson99,yanson00,yanson01a}. Peaks have been found to
persist up to conductances $\simeq 100 G_0$, while the
peak positions are periodic in $\sqrt{G}$.  This is evidence for
shell-filling effects, comparable to what happens in metal clusters.  The
amplitude of the peaks has been found to be further periodically modulated,
reflecting a supershell structure\cite{yanson00}.  The same type of shell
effects have recently been observed for gold
nanowires\cite{diaz03,mares-hulea03}.  These experiments, though not
directly accessing the lifetime of nanowires, still provide some
information about it: A nanowire might be stable to small perturbations,
but not give rise to a peak in a conductance histogram provided its lifetime is short compared to the rate of deformation of the contact in the experiment. The observation
of a given peak therefore gives access to a lower bound of the lifetime of
the corresponding wire.

Imaging experiments\cite{kondo97,ohnishi98,okamoto99,kondo00,rodrigues00,oshima03a}
using TEM give a more direct access to the lifetime of nanowires, and
produce long cylindrical wires which are more easily treated
theoretically. Holes are burned through a thin gold film using an intense
electron beam, leaving a suspended nanowire when two holes come close
together. Such a wire is often found to evolve into a long, nearly perfect
cylindrical wire, which connects to macroscopic contacts at two well-defined
junctions.\cite{kondo97}  Under electron-beam irradiation, the wires
are observed to thin via the nucleation of a surface kink at one end, and
the subsequent propagation of the kink along the wire until it is absorbed
in the other contact.
Although no systematic study of the lifetime of such
wires is available, they have been reported to remain stable for
seconds\cite{kondo97,kondo00} despite the strong electron irradiation necessary for the imaging
process.

\section{THE MODEL}
\label{sec:model}

We use a nanoscale free-electron model\cite{stafford97a} in which the
atomic structure of the wire is replaced by a continuous positive
background of constant charge density. We restrict ourselves to wires with
axial symmetry (which includes the most stable wires\cite{Urban04}), 
described by the radius of the wire
$R(z,t)$ at time $t$ and position $z$ along the wire axis.  
Free and independent electrons
are confined within the wire by hard-walls.  The nanowire is in electrical
contact with macroscopic metallic electrodes,\cite{kondo97,kondo00} so 
the relevant thermodynamic potential for the electrons is the
grand-canonical potential
\begin{equation}\label{eq:omega}
\Omega_e(T,\mu_e) = -k_B
T\int\text{d}E\,g(E)\ln\left(1+e^{-\frac{E-\mu_e}{k_BT}}\right)
\end{equation}
at temperature $T$ and electrochemical potential $\mu_e$,
$k_B$ being the Boltzmann constant (which will hereafter be set to one).
An expansion of $\Omega$ in terms of geometrical quantities can generically be
written\cite{brack97,Stafford99} as
\begin{equation}
\label{eq:weyl}
\Omega_e = -\omega{\cal V} + \sigma{\cal S} - \gamma{\cal C} + \delta\Omega,
\end{equation}
where ${\cal V, S, C}$ are respectively the volume, surface area and
integrated mean curvature of the wire, and $\omega, \sigma, \gamma$ are
material and temperature dependent coefficients. $\delta\Omega$ is a
fluctuating term giving quantum corrections to the otherwise smooth
expansion of $\Omega$.

\begin{figure}[ht]
   \centering
   \includegraphics[width=10cm,angle=0]{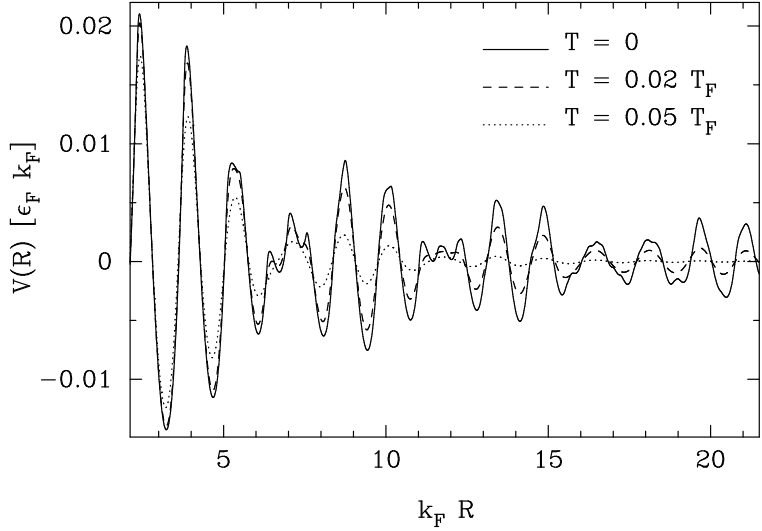} 
   \caption{Electron-shell potential $V_{shell}(R,T)$ at zero and two finite
temperatures, which correspond respectively to 1000K and 2500K for Na.
The electrical conductance values of the principal wires studied in this article
are indicated on the upper horizontal axis.}
   \label{fig:potential}
\end{figure}

Assuming the radius of the wire changes slowly compared to the
Fermi wavelength $\lambda_F$, the fluctuating term $\delta\Omega$ can be
written as
\begin{equation}
\label{eq:deltaOm}
\delta\Omega = \int\text{d}z\; V_{shell}(R(z),T),
\end{equation}
where $V_{shell}(R,T)$ is the electron-shell potential, shown in
Fig.~\ref{fig:potential} for three different temperatures.
This potential is responsible for the stability of metallic nanowires, its
minima corresponding to the ``magic'' radii mentioned above, and can be
computed in the semiclassical approximation using a Gutzwiller-type trace
formula as\cite{burki03}
\begin{equation}\label{eq.gutzwiller}  
V_{shell}(R,T) = \frac{2\varepsilon_F}{\pi} \sum_{w=1}^{\infty}\sum_{v=2w}^{\infty}
a_{vw}(T) \frac{f_{vw}\cos\theta_{vw}}{v^2L_{vw}},
\end{equation}  
where the sum includes all classical periodic orbits $(v,w)$ in a disk
billiard \cite{brack97}, characterized by their number of vertices $v$ and
winding number $w$, $L_{vw}=2vR\sin(\pi w/v)$ is the length of orbit
$(v,w)$, and $\theta_{vw}=k_FL_{vw}-3v\pi/2$.  The factor $f_{vw}=1$
for $v=2w$, $f_{vw}=2$ otherwise,
accounts for the invariance under time-reversal
symmetry of some orbits, and $a_{vw}(T) = \tau_{vw}/\sinh{\tau_{vw}}$
($\tau_{vw}=\pi k_FL_{vw}T/2T_F$) is a temperature-dependent damping
factor.

Such a model has been successful in explaining many of the observed
phenomena in monovalent metallic nanowires: sawtooth behavior of the
tensile force, correlated with conductance steps\cite{stafford97a};
conductance histograms\cite{burki99}; spontaneous formation\cite{burki03}
and stability\cite{kassubek00,zhang02a,urban03} of long cylindrical
nanowires; shell and super-shell structure in conductance
histograms\cite{Urban04}, etc.  The latter study relaxed the restriction to
axial symmetry, showing that the most stable wires are indeed cylindrical,
with a few exceptions at low conductance, where a few additional stable
wires with elliptic cross-sections have been found. One may conclude that
deformations breaking the axial symmetry can essentially be ignored for
thicker wires, due to their large surface energy cost.

In the spirit of the Born-Oppenheimer approximation, the electronic energy
functional (\ref{eq:weyl}) is taken as the potential energy of the ionic
medium.  The nanowire is in contact with metallic electrodes, with which it
can exchange atoms via surface self-diffusion.\cite{burki03}  The relevant
thermodynamic potential for the atoms is thus
\begin{equation}
\Omega_a = \Omega_e - \mu_a {\cal V}/{\cal V}_a,
\label{eq:energy_atoms}
\end{equation}
where $\mu_a$ is the chemical potential for a surface atom in the contacts, and
${\cal V}_a =3\pi^2/k_F^3$ is the volume of an atom.
Eq.\ (\ref{eq:energy_atoms}) is a functional of the geometry of the contact, 
and is the basis for our field-theory.
The term proportional to the mean curvature in Eq.\ (\ref{eq:weyl}) has a
contribution from the transverse curvature, which is proportional to the
length $L$ of the wire, and a contribution from the longitudinal curvature,
which is only a few percent of the surface energy, and can therefore be
neglected.
As a result, the energy (\ref{eq:energy_atoms}) becomes
\begin{equation}
\label{eq:omega2}
\Omega_a
= \int\text{d}z\left[2\pi\sigma
R(z,t)\sqrt{1+(\partial_zR)^2} -\pi\gamma + V_{shell}(R,T) \right]
-(\omega + \mu_a/{\cal V}_a){\cal V},
\end{equation}
where $\partial_zR$ is the derivative of $R(z,t)$ with respect to $z$.

We are considering the lifetime of perfect cylinders of length $L$, so that the 
radius function may be written as a constant plus fluctuations,
\begin{equation}
\label{eq:radius}
R(z,t) \equiv R_0 + \phi(z,t).
\end{equation}
A stable cylindrical nanowire of radius $R_0$ represents a state of diffusive 
equilibrium between the wire and the contacts, for which
\begin{equation}
\frac{\mu_a}{{\cal V}_a} = \frac{1}{2\pi R_0}
\frac{\partial}{\partial R_0} \left(\frac{\Omega_e(R_0)}{L}\right)
= \frac{\sigma}{R_0} - \omega
+ \frac{1}{2\pi R_0} \frac{\partial V_{shell}(R_0)}{\partial R_0},
\label{eq:mu_a}
\end{equation}
where $\Omega_e(R_0)$ is the electronic energy of an unperturbed cylinder.
The most stable nanowires correspond to minima of $V_{shell}(R_0)$ 
(c.f.\ Fig.\ \ref{fig:potential}), for which Eq.\ (\ref{eq:mu_a}) simplifies to
\begin{equation}
\frac{\mu_a}{{\cal V}_a} = \frac{\sigma}{R_0} - \omega.
\label{eq:mu_a_stable}
\end{equation}
The energy (\ref{eq:omega2}) can be expanded in a series in
$\phi$. Inserting Eq.\ (\ref{eq:mu_a_stable}) into Eq.\ (\ref{eq:omega2}),
one sees that the term linear in $\phi$ vanishes.
Ignoring higher order terms in $\partial_z \phi$, one gets
$\Omega_a = \Omega_a(R_0) + {\cal H}[\phi]$, where $\Omega_a(R_0)$ is the
energy of an unperturbed cylinder and
\begin{equation}
\label{eq:energy}
{\cal H}[\phi] = \int_0^L\!\!\text{d}z\left[\frac{\kappa}{2}(\partial_z\phi)^2
+ V(\phi)\right].
\end{equation}
Here $\kappa=2\pi\sigma R_0$ and
\begin{equation}
\label{eq:Vtilde}
V(\phi) \equiv V_{shell}(R_0+\phi) - V_{shell}(R_0)-\frac{\pi\sigma}{R_0}\phi^2.
\end{equation}

Several minima in the potential $V(\phi)$, such as
those corresponding to stable wires with conductance 
$G/G_0=3,6,17,23,42,51,96,\dots$ (cf.\ Fig.\ \ref{fig:potential})
can be locally approximated either by a bistable symmetric quartic potential
\begin{equation}
\label{eq:quartic}
V^{s}(\phi) = \frac{1}{2}\eta \phi^2 - \frac{1}{4}\lambda \phi^4,
\end{equation}
or by an asymmetric cubic potential
\begin{equation}
\label{eq:cubic}
V^{a}(\phi) = -\alpha \tilde\phi + \frac{1}{3}\beta \tilde\phi^3\, ,
\end{equation}
where $\tilde\phi=\phi+\sqrt{\alpha/\beta}$.  The latter potential biases fluctuations towards
smaller radii ($\phi<0$); however biases in the opposite
direction are trivially accommodated by reflecting the potential so that
$\tilde\phi\mapsto -\tilde\phi$ and redefining $\tilde\phi=\phi-\sqrt{\alpha/\beta}$.

The boundary conditions on $\phi$ are determined by
the physics of the problem:  Simulations of nanowire surface
dynamics\cite{burki03} indicate
that the cylindrical segment of a nanowire joins
abruptly to a contact having the form of an unduloid of revolution (for
which the electron-shell correction to the energy is suppressed by
a finite slope), consistent with Neumann boundary conditions.
This choice of boundary conditions is also consistent with the experimental
finding\cite{kondo97} that thinning of a suspended nanowire occurs via
nucleation of a surface kink at one end, as discussed in Sec.\ \ref{sec:states}.

Finally, we note that in this paper, we consider a thermodynamic ensemble of nanowires at
{\em fixed length} $L$.  This implies that the ends of the wire are
held fixed by a tensile force $F_{\rm eq}=-\partial
\Omega_e/\partial L$, which was calculated
previously.\cite{stafford97a,zhang02a,Stafford99}
Under elongation or compression at a finite rate, $F\neq F_{\rm eq}$, and
hence a different thermodynamic ensemble must be used.  The dependence of
nanowire lifetime on the applied force will be discussed
elsewhere.\cite{burki04b}

Eqs.~(\ref{eq:energy}-\ref{eq:cubic}) provide the starting point for our
field-theoretic formulation of nanowire lifetimes, to be developed in the
following sections.

\section{Escape Phenomenology}
\label{sec:escape}

The preceding discussion suggests that the problem of stability of
nanowires against thermal fluctuations can be studied as a one-dimensional
Ginzburg--Landau scalar field theory, perturbed by weak spatiotemporal
noise, in a domain of finite extent.  Problems of this type have recently
received increasing attention.  The classical nucleation problem on an {\it
infinite\/} line was treated by Langer~\cite{Langer69}, and its quantum
analogue by Callan and Coleman~\cite{CC77}, in two influential and
much-cited papers.  However, the corresponding problem in a {\it finite\/}
domain has been less intensively studied~\cite{FJ82,MOS89,MT95}.

The difference between finite and infinite systems is not merely
quantitative.  For example, it was recently found that an unusual effect,
analogous to a phase transition, occurs in an overdamped classical
Ginzburg--Landau field theory with a bistable $\phi^4$ potential.
The transition occurs when the length~$L$ of its one-dimensional spatial domain is
varied~\cite{MS01,MS03}.  Below a critical length~$L_c$, the transition
state is a spatially constant field configuration.  However, at~$L=L_c$ it
bifurcates in the symmetric case into a spatially varying pair of
configurations, degenerate in energy.  The asymmetric potential of
Eq.~(\ref{eq:cubic}) shows a similar transition~\cite{St04}, but due to the
asymmetry there is only a single preferred activation state.  These studies
demonstrate that the ``phase transition'' at a critical length $L_c$ is
reasonably robust.

As~one would expect, the transition rate is strongly affected by the
transition.  Formally, the prefactor in the Kramers (weak-noise) nucleation
rate {\em diverges\/} at~$L=L_c$.  This signals that precisely at $L=L_c$,
escape from a stable state becomes {\it non-Arrhenius\/}: the rate at which
it occurs falls off in the limit of weak noise not like an exponential
(with a constant prefactor), but rather like an exponential with a
power-law prefactor.
An interesting consequence is that this transition may be observable in
nanowire decay phenomenology.

The model as introduced in Sec.~\ref{sec:model} treats $\phi$, the
fluctuations of the nanowire radius, as a classical field on a
one-dimensional spatial domain $[0,L]$.  Its dynamics are governed by the
stochastic Ginzburg--Landau equation
\begin{equation}
\label{eq:GL}
\dot\phi=\kappa\phi'' -V'(\phi) + (2T)^{1/2}\xi(z,t),
\end{equation}
where $\xi(z,t)$ is unit-strength
spatiotemporal white noise, satisfying
$\langle\xi(z_1,t_1)\xi(z_2,t_2)\rangle=\delta(z_1-z_2)\delta(t_1-t_2)$.
In Eq.\ (\ref{eq:GL}), time is measured in units of a microscopic
timescale describing the short-wavelength cutoff of the surface
dynamics,\cite{zhang02a} which is given to within a factor of order
unity by the inverse Debye frequency $\nu_D^{-1}$.
The zero-noise dynamics is ``gradient,'' i.e., conservative.  That~is,
at zero temperature
\begin{equation}
\label{eq:gradient}
\dot\phi=-\delta{\cal H}/\delta\phi\, ,
\end{equation}
where ${\cal H}[\phi]$ is the energy functional, given by Eq.\ (\ref{eq:energy}).
So the statistical properties of the
stochastically evolving field~$\phi$ are described by equilibrium
statistical mechanics.  At nonzero temperature, however, thermal
fluctuations can induce transitions between stable states (i.e., local
minima) of the potential $V(\phi)$, which correspond in our model to
different stable cylindrical radii, as discussed in Sec.~\ref{sec:model}.  Such
transitions occur via nucleation of a ``droplet'' of one stable configuration
in the background of the other, subsequently quickly spreading to fill the
entire spatial domain.  When the noise is weak, i.e., at low temperatures
(compared to the barrier height) most fluctuations will not succeed in
nucleating a new phase; it is far more likely for a small droplet to shrink
and vanish.

In the infinite-dimensional configuration space, a transition state must go
``uphill'' in energy from each stable field configuration.  Because of
exponential suppression of fluctuations as their energy increases, there is
at low temperature a preferred transition configuration (saddle) that lies
between adjacent minima.  These are the nucleation pathways, in effect
``paths of least resistance.''  By time-reversal invariance, they are
time-reversed zero-noise ``downhill'' trajectories~\cite{MS93}.  At low
temperatures, the expected waiting time of the order parameter $\phi$ in a
basin of attraction is an exponential random variable, as is typical of
slow rate processes.  The activation rate (the reciprocal of the mean time
between flips) will be given in the $T\to0$ limit by the Kramers formula
\begin{equation}
\label{eq:Kramers}
\Gamma\sim \Gamma_0\exp(-\Delta E/T)\, .
\end{equation}
Here $\Delta E$ is the activation barrier, which quantifies the extent to
which the preferred transition configuration between the two stable
configurations is energetically disfavored; that is, $\Delta E$ is the
energy of the transition state minus the energy of either stable state.
$\Gamma_0$ is the rate prefactor. 

The quantities $\Delta E$ and $\Gamma_0$ depend on the parameters of the
potential, on the length~$L$, and on the choice of boundary conditions at
the endpoints $z=0$ and~$z=L$.  The boundary conditions affect the way in
which order parameter reversal occurs, since they may force nucleation to
begin, preferentially, at the endpoints.  

\section{The Stable and Transition States}
\label{sec:states}

It will simplify the discussion to express the potentials in
Eqs.~(\ref{eq:quartic}) and (\ref{eq:cubic}) in dimensionless units.

For the symmetric potential Eq.~(\ref{eq:quartic}), we can scale out the
various constants by introducing the variables $x=\sqrt{\eta/\kappa}z$,
$u=\sqrt{\lambda/\eta}\phi$, and $E_0=\kappa^{1/2}\eta^{3/2}/\lambda$.
The energy functional then becomes
\begin{equation}
\label{eq:symscaled}
{\cal H}[u]/E_0=\int_0^{\ell^s} \left[\frac{1}{2}(u')^2 +\frac{1}{2}u^2 -
\frac{1}{4}u^4\right] \,dx
\end{equation}
where $\ell^s=\sqrt{\eta/\kappa}L$.

\begin{figure}[ht]
\begin{center}
\begin{tabular}{c}
\includegraphics[height=5cm, width=12cm]{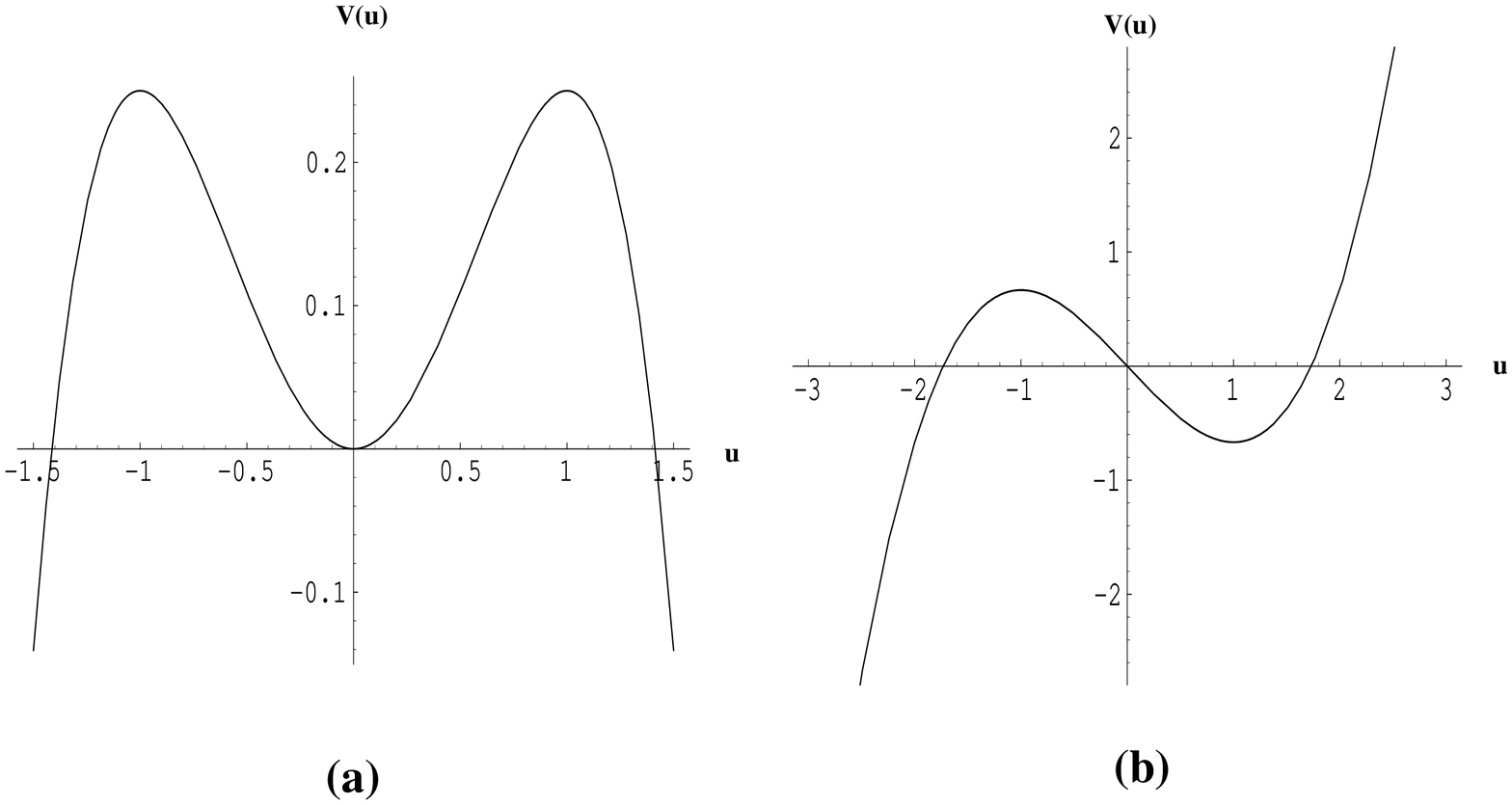} 
\end{tabular}
\end{center}
\caption[The potential.]{Potentials for the reduced order parameter~$u$.
(a) Symmetric case corresponding to Eq.~(\ref{eq:quartic}).  (b) Asymmetric
case corresponding to Eq.~(\ref{eq:cubic}).}
\label{fig:potentials}
\end{figure}

For the asymmetric potential Eq.~(\ref{eq:cubic}), we can scale out the
various constants by introducing the variables
$x=\Big[(\alpha\beta)^{1/4}/\kappa^{1/2}\Big]z$,
$u=\sqrt{\beta/\alpha}\tilde\phi$, and
$E_0=\kappa^{1/2}\alpha^{5/4}/\beta^{3/4}$.  The energy functional then
becomes
\begin{equation}
\label{eq:asymscaled}
{\cal H}[u]/E_0=\int_{0}^{\ell^a} \left[\frac{1}{2}(u')^2  -u + \frac{1}{3}u^3\right]\,dx
\end{equation}
where $\ell^a=\Big[(\alpha\beta)^{1/4}/\kappa^{1/2}\Big]L$.
These reduced potentials are shown in Fig.~\ref{fig:potentials}.

In the absence of external noise, and with Neumann boundary conditions
$u'(0)=u'(L)=0$, the noiseless ($T=0$) evolution equation
(\ref{eq:GL}) with either potential possesses both uniform and
nonuniform stationary states.  We turn now to their study.

\subsection{Symmetric Potential}
\label{subsec:symmetric}

\begin{figure}[hb]
\begin{center}
\begin{tabular}{c}
\includegraphics[height=6.25cm]{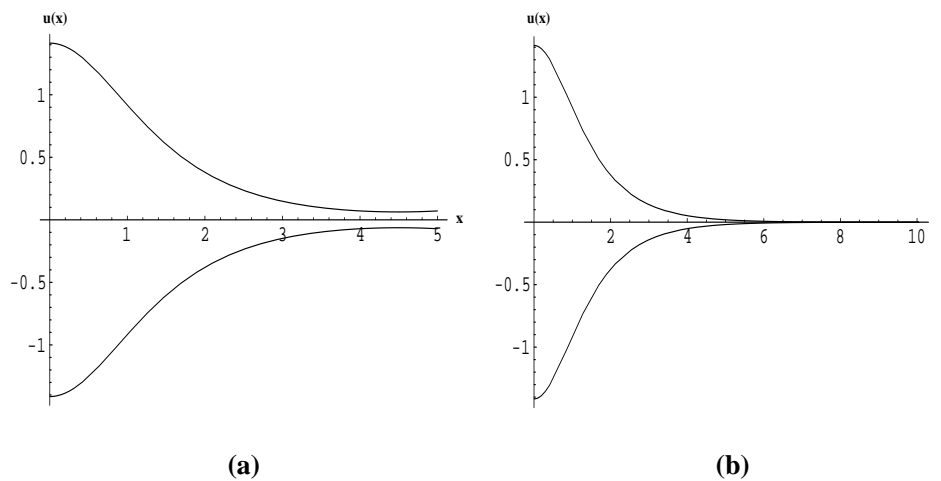} 
\end{tabular}
\end{center}
\caption[The stationary states.]{The two lowest-energy nonconstant
transition states for the symmetric potential with Neumann boundary
conditions, when (a) $\ell^s=5$ and (b) $\ell^s=10$.  Because of the
symmetry, each nonzero state has a degenerate counterpart obtained by
$u\mapsto -u$.  The width of the nonconstant portion of the instanton
remains essentially constant for larger $\ell^s$.  The stable state for any
$\ell^s$ is $u=0$.}
\label{fig:symsol}
\end{figure}

There are three constant time-independent solutions: $u=0$, with energy 0,
and $u=\pm 1$, with energy~$1/4$.  It is easy to see from
Fig.~\ref{fig:potentials} that $\phi=0$ is stable for any~$L$, and
$\phi=\pm 1$ are always unstable; this is confirmed by eigenvalue
analysis~\cite{St04} (see below).

What {\em nonconstant\/} time-independent solutions exist when $T=0$?
By Eqs.\ (\ref{eq:energy}) and (\ref{eq:gradient}), any stationary solution
satisfies $\dot\phi=-\delta{\cal H}/\delta\phi=0$, yielding in this case
the nonlinear ordinary differential equation
\begin{equation}
\label{eq:ELsym}
u''=u-u^3\, .
\end{equation}
The linearized noiseless dynamics in the vicinity of such a state are
specified by the Hessian operator $\delta^2{\cal H}/\delta u^2$.  Stability
of stationary solutions is determined by eigenvalues of this operator.
Those with all positive eigenvalues are stable; those with a single
negative eigenvalue are potential transition states.  (In the limit of low
thermal noise, the actual transition state is that with the smallest energy
difference with the relevant stable state.) The eigenvector corresponding
to the negative eigenvalue is the direction (in the infinite-dimensional
configuration space) along which the optimal escape trajectory approaches
the saddle, in the limit of low noise.

Kink-like field configurations that asymptotically connect ground or vacuum
states are often called {\em instanton states\/}, in a nomenclature derived
from Callan and Coleman~\cite{CC77}.  We will use the term ``instanton'' here
for similar nonconstant solutions on finite domains.

When $\ell^s$ is finite, the instanton state(s) can be expressed in terms
of {\it Jacobi elliptic functions\/}~\cite{Abramowitz65}.  Such functions
are characterized by an index $m$, $0\leqslant m\leqslant 1$, which we shall see is
related to $\ell^s$ through the boundary conditions.  The instanton solution
of Eq.~({\ref{eq:ELsym}) is the same as that for periodic boundary conditions~\cite{MT95}, but the dependence of $\ell^s$ on $m$ differs:
\begin{equation}
\label{eq:syminstanton}
u_{{\rm inst},m}(x) = \pm\sqrt{\frac{2}{2-m}}\,{\rm
dn}(x/\sqrt{2-m}\mid m),
\end{equation}
where ${\rm dn}(\cdot\mid m)$ is the Jacobi elliptic $\rm dn$ function with
parameter~$m$.  Its half-period is given by~${\bf K}(m)$, the complete
elliptic integral of the first kind~\cite{Abramowitz65}, which is a
monotonically increasing function of~$m$.  From a physical perspective such
a solution, extended over the whole line, can be viewed as an infinite
alternating sequence of kinks and anti-kinks, spaced a distance ${\bf
K}(m)$ apart.  As~$m\to 0^+$, ${\bf K}(m)$ decreases to~$\pi/2$, and ${\rm
dn}(\cdot\mid m)$ degenerates to 1.  As $m\to 1^-$, the half-period ${\rm
dn}(\cdot\mid m)$ increases to infinity (with a logarithmic divergence),
and ${\rm dn}(\cdot\mid m)$ degenerates to the nonperiodic function
sech$(\cdot)$, which is the canonical double-kink soliton.  This limiting
form is in fact the shape of the critical droplet pair in the Langer and
Callan--Coleman analyses.  (A good pedagogical discussion is given by Schulman~\cite{Schulman81}).

It is easily seen that the solution Eq.~(\ref{eq:syminstanton}) with lowest
energy (i.e., fewest kinks) that satisfies the Neumann boundary condition
requires
\begin{equation}
\label{eq:sybc}
\ell^s=\sqrt{2-m}{\bf K}(m)\, ,
\end{equation}
which in turn leads to $\ell^s_c=\pi/\sqrt{2}$.  As $\ell^s\to\ell^s_c$
from above (i.e., $m\to 0^+$), ${\rm dn}(x|0)=1$, and the instanton states
reduce to the uniform unstable states $u_u=\pm1$.  This point corresponds
to the {\it critical length\/} $\ell^s_c$~\cite{MS01,MS03}.
If $\ell^s\leqslant\ell^s_c$ the transition state
is one of the two uniform configurations $u=\pm 1$.  As $m\to 1^-$,
$\ell\to\infty$, and the instanton state becomes
\begin{equation}
\label{eq:sech}
u_{{\rm inst},1}(x)=\sqrt{2}{\rm sech}(x)\,
\end{equation}

The nonconstant transition states for the symmetric potential case with
Neumann boundary conditions are plotted in Fig.~\ref{fig:symsol}. Of course,
by the symmetry in the problem the instanton can nucleate with equal
probability at either end; this is manifested by the existence of
degenerate solutions (not shown) in which $x$ is replaced by $L-x$ in
Eq.~(\ref{eq:syminstanton}).  This will be true for the asymmetric case as
well, and the existence of solutions reflected about the midpoint of the
interval will be tacitly understood from now on.

\subsection{Asymmetric Potential}
\label{subsec:asymmetric}

\begin{figure}[b]
\begin{center}
\begin{tabular}{c}
\includegraphics[height=6.25cm]{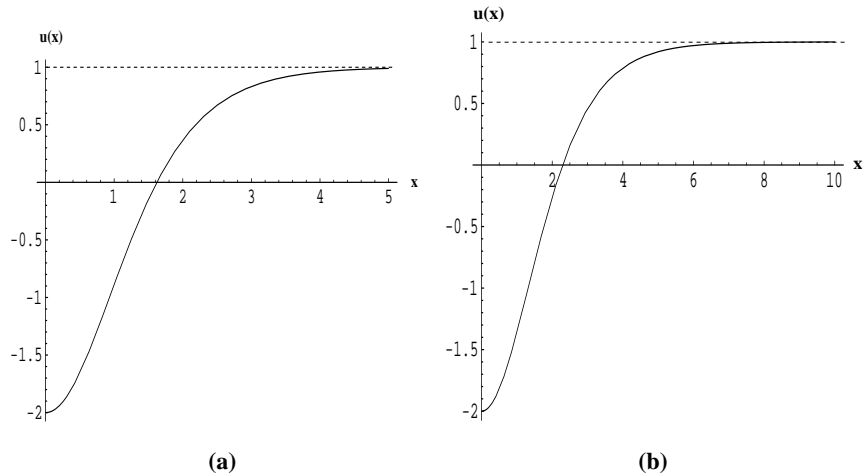} 
\end{tabular}
\end{center}
\caption[The stationary states.]{The stable state (dashed line) and
nonconstant transition state (solid line), for the asymmetric potential
with Neumann boundary conditions, when (a) $\ell^a=5$ and (b)
$\ell^a=10$.
The width of the nonconstant portion of the instanton
remains essentially constant for larger $\ell^a$.}
\label{fig:asymsols}
\end{figure}

Here the time-independent solutions of the zero-noise Ginzburg-Landau
equation (i.e., extremal states of ${\cal H}[\phi]$) satisfy
\begin{equation}
\label{eq:extrema}
u''=-1+u^2\, .
\end{equation}

With Neumann boundary conditions, there is a uniform stable state $u_s=+1$,
and a uniform unstable state $u_u=-1$.  The latter is the transition state
for $\ell^a<\ell^a_c=\pi/\sqrt{2}$.  At $\ell^a_c$ a transition occurs, and
above it the transition state is nonuniform.

In the asymmetric case the instanton solution is~\cite{St04}
\begin{equation}
\label{eq:instanton}
u_{{\rm inst},m}(x) = \frac{2-m}{\sqrt{m^2-m+1}} -
\frac{3}{\sqrt{m^2-m+1}}{\rm dn}^2\left(\frac{x}{\sqrt{2}(m^2-m+1)^{1/4}} \Big| m\right)\, .
\end{equation}
It is easily seen that this solution satisfies the Neumann boundary
condition with lowest energy when
\begin{equation}
\label{eq:asybc}
\ell^a=\sqrt{2}(m^2-m+1)^{1/4}{\bf K}(m)\, ,
\end{equation}
which in turn leads to $\ell^a_c=\pi/\sqrt{2}$, as noted above.  (The fact
that the critical lengths in the symmetric and asymmetric case are
equal is coincidental.)  As $\ell^a\to\ell^a_c$ from above (i.e., $m\to
0^+$), ${\rm dn}(x|0)=1$, and the instanton state reduces to the uniform
unstable state $u_u=-1$.  As $m\to 1^-$, $\ell\to\infty$, and the instanton
state becomes
\begin{equation}
\label{eq:asech}
u_{{\rm inst},1}(x)=1-3{\rm sech}^2\left(\frac{x}{\sqrt{2}}\right)\,
\end{equation}

The stable and transition states for the asymmetric potential case with
Neumann boundary conditions are shown in Fig.~\ref{fig:asymsols}.

\section{The Activation Barrier}
\label{sec:action}

As mentioned in Sec.~\ref{sec:escape}, the exponential falloff of the
transition rate in the limit of weak thermal noise, i.e., its Arrhenius
behavior, is determined by the activation barrier (sometimes called the
``activation energy'') $\Delta E$ between the stable and instanton states.
$\Delta E$ is defined to be $({\cal H}[\phi_u]-{\cal H}[\phi_s])$, with the
energy functional ${\cal H}[\phi]$ given by Eq.\ (\ref{eq:energy}).  The
calculation of ${\cal H}[\phi_s]$ and ${\cal H}[\phi_u]$, the stable and
transition state energies, is trivial in the case of uniform states.  It
remains reasonably straightforward in the case of the instanton states for
the models studied here, and can be expressed in terms of complete elliptic
integrals~\cite{Abramowitz65}.  The results are as follows:

\medskip

\noindent {\em Symmetric potential.\/}---If $\ell^s\leqslant
\ell^s_c=\pi/\sqrt{2}$, then the transition state is one of the two uniform
configurations $u=\pm 1$, and $\Delta E/E_0 = \ell^s/4$.  If $\ell^s>\ell^s_c$,
then
\begin{equation}
\label{eq:actn}
\Delta E^s/E_0 = {1\over 3\sqrt{2-m}}\left[2{\bf E}(m)
-{1-m\over 2-m}{\bf K}(m)\right]\, .
\end{equation}
The activation energy in the $\ell^s\to\infty$ limit equals $2/3$, which is
the energy of a single kink.

\medskip

The above formula for the activation energy $\Delta E^s$ as a function of
$\ell^s$ is plotted in Fig.~\ref{fig:symacten}.  The transition at
$\ell^s_c=\pi/\sqrt{2}$ is apparent, as is the differentiability
(and lack of twice differentiability) through the transition.

\begin{figure}[ht]
\begin{center}
\begin{tabular}{c}
\includegraphics[height=6.25cm]{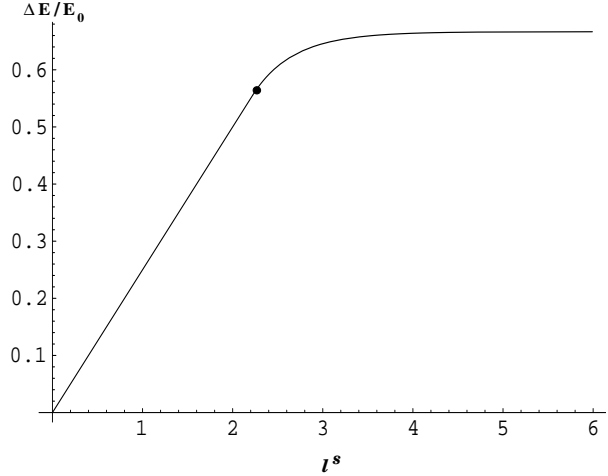} 
\end{tabular}
\end{center}
\caption[The activation energy.]{The activation energy $\Delta E^s$ as a
function of the interval length~$\ell$, for the symmetric potential with
Neumann boundary conditions.  The bullet indicates the critical interval
length $\ell^s_c=\pi/\sqrt{2}$ at which the transition takes place.}
\label{fig:symacten}
\end{figure}

\medskip

\noindent {\em Asymmetric potential.\/}---If $\ell^a\leqslant
\ell^a_c=\pi/\sqrt{2}$, then the transition state is the uniform
configuration $u=-1$, and $\Delta E^a/E_0 = (4/3)\ell^a$.  If $\ell^a>\ell^a_c$,
then
\begin{equation}
\label{eq:barrier}
{\Delta E^a\over E_0} = 
\left[{2-3m-3m^2+2m^3\over 3(m^2-m+1)^{3/2}}+{2\over 3}\right]\ell^a
+{6\sqrt{2}\over 5(m^2-m+1)^{1/4}}
\left[2{\bf E}(m)-{(2-m)(1-m)\over(m^2-m+1)}{\bf K}(m)\right]\, .
\end{equation}
There is a difference of a factor of 2 in the second term of the RHS with
respect to the corresponding formula in Ref.\ \cite{St04};
this results from the
different boundary condition employed in that paper. As $\ell\to\infty$,
$\Delta E^a/E_0\to 12\sqrt{2}/5$.  The activation barrier for the entire
range of $\ell$ is shown in Fig.~\ref{fig:asymacten}.

\begin{figure}[t]
\begin{center}
\begin{tabular}{c}
\includegraphics[height=6.25cm]{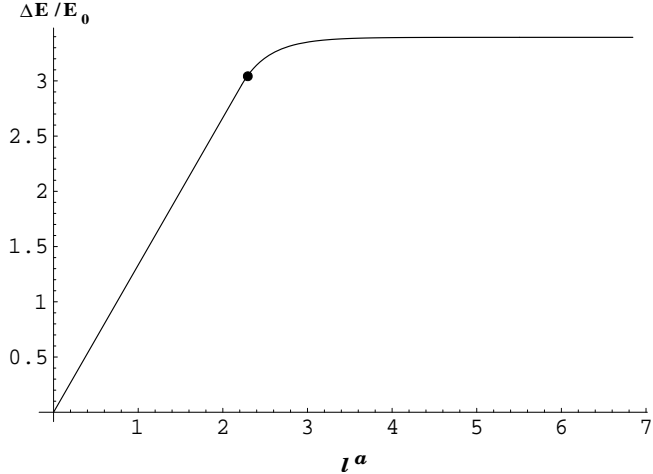} 
\end{tabular}
\end{center}
\caption[The activation energy.]{The activation energy $\Delta E^a$ as a
function of the interval length~$\ell^a$, for the asymmetric potential with
Neumann boundary conditions.  The bullet indicates the critical interval
length $\ell^a_c=\pi/\sqrt{2}$ at which the transition takes place.}
\label{fig:asymacten}
\end{figure}

\section{The Transition Rate Prefactor}
\label{sec:prefactor}

Calculation of the prefactor $\Gamma_0$ in the Kramers transition rate
formula (\ref{eq:Kramers}) is a much more involved matter than the computation of $\Delta E$.  It generally requires an analysis of the {\em transverse fluctuations\/} about the instanton
solutions.  The general method applied to the present set of problems has been 
discussed elsewhere~\cite{MS01,MS03,St04}.  In the present article, we
focus on the most technologically relevant regime $L\gg L_c$, in which the 
prefactor is $O(1)$ in natural units, i.e., $\Gamma_0 \approx \nu_D$.
We defer discussion of the prefactor in shorter wires with $L \lesssim L_c$
to a later publication.\cite{burki04b}

\section{DISCUSSION}
\label{sec:discussion}

Pulling together the results from the preceding sections, our continuum
dynamical model predicts that the lifetime $\tau$ of a metastable
cylindrical nanowire of length greater than the critical length $L_c$ is
given by
\begin{equation}
\frac{1}{\tau} = \Gamma \approx  \nu_D \exp(-\Delta E_\infty/T),
\label{eq:tau}
\end{equation}
where $\nu_D$ is the Debye frequency and $\Delta E_\infty =
\lim_{L\rightarrow \infty} \Delta E$ is the activation barrier for a long
wire.  Note that for a moderately thick wire with $G/G_0\gg 1$, the lifetime
$\tau$ may not be the typical time before the wire breaks, but rather a
switching time between two different metastable wires with different
conductance values.
In terms of the physical parameters defined in section \ref{sec:model},
$\Delta E_\infty=\frac{2}{3} \kappa^{1/2} \eta^{3/2}/\lambda$ for the case of a symmetric
quartic potential well (\ref{eq:quartic}) and
$\Delta E_\infty=\frac{12\sqrt{2}}{5} \kappa^{1/2} \alpha^{5/4}/\beta^{3/4}$ for the case
of an asymmetric cubic potential well (\ref{eq:cubic}).  The lifetimes for
several cylindrical sodium and gold nanowires, calculated using the best
cubic- or symmetric quartic-polynomial fits to the potential
(\ref{eq:Vtilde}), are tabulated in Table \ref{tab:lifetime}.

\begin{table}[ht]
\caption{The lifetime $\tau$ (in seconds)
for various cylindrical sodium and gold
nanowires at temperatures from 75K to 200K.
Here $G$ is the electrical conductance
of the wire, $L_c$ is the critical length above which the lifetime may
be approximated by $\tau \approx \nu_D^{-1} \exp(\Delta E_\infty/T)$,
$\Delta E_\infty$ is the activation energy for an infinitely long wire,
and $\nu_D$ is the Debye frequency.  Note that wires shorter than $L_c$
are predicted to have shorter lifetimes.
}
\label{tab:lifetime}
\begin{center}
\begin{tabular}{|c||c|c|c|c|c||c|c|c|c|c|}
\hline
\rule[-1.5ex]{0pt}{4.5ex} & \multicolumn{5}{c||}{Na} & \multicolumn{5}{c|}{Au} \\
\cline{2-11}
\rule[-1.5ex]{0pt}{4.5ex} $G$ & $L_c $ & $\Delta E_\infty $ &  \multicolumn{3}{c||}{$\tau$ [s]}  & $L_c $ & $\Delta E_\infty $ &  \multicolumn{3}{c|}{$\tau$ [s]} \\
\cline{4-6}\cline{9-11}
\rule[-1.5ex]{0pt}{4.5ex} [$G_0$] & [nm] & [meV] & $75\;$K & $100\;$K & $150\;$K & [nm] & [meV] & $100\;$K & $150\;$K & $200\;$K \\
\hline\hline
\rule[0ex]{0pt}{3.ex}%
3 & 1.4 &  210 & 30 &  $9\times10^{-3}$ &  $3\times10^{-6}$ & 1.5 &  470 &  $10^{11}$ &  $2\times10^{3}$ & 0.2 \\
6 & 2.6 &  170 & 0.06 &  $9\times10^{-5}$ &  $10^{-7}$ & 3.0 &  310 &  $10^{3}$ &  $7\times10^{-3}$ &  $2\times10^{-5}$ \\
17 & 3.1 &  230 &   500 & 0.08 &  $10^{-5}$ & 3.4 &  470 &  $9\times10^{10}$ &  $10^{3}$ & 0.2 \\
23 & 3.7 &  190 & 4 &  $2\times10^{-3}$ &  $10^{-6}$ & 4.1 &  390 &  $10^{7}$ & 3 &  $2\times10^{-3}$ \\
42 & 4.3 &  210 & 50 & 0.01 &  $4\times10^{-6}$ & 4.8 &  440 &  $3\times10^{9}$ &   100 & 0.03 \\
51 & 4.5 &  150 &  $7\times10^{-3}$ &  $2\times10^{-5}$ &  $5\times10^{-8}$ & 4.9 &  320 &  $2\times10^{3}$ & 0.01 &  $3\times10^{-5}$ \\
\rule[-1.5ex]{0pt}{3.ex}%
96 & 5.8 &  200 & 5 &  $2\times10^{-3}$ &  $10^{-6}$ & 6.3 &  440 &  $8\times10^{9}$ &   300 & 0.05 \\
\hline
\end{tabular}
\end{center}
\end{table}

An important prediction given in Table \ref{tab:lifetime} is that the 
lifetimes of the most stable nanowires, while they do exhibit significant 
variations from one conductance plateau to another, do not vary systematically
as a function of radius; the activation barriers in Table \ref{tab:lifetime}
vary by only about 
30\% from one plateau to another, and the wire with a conductance of
$96 G_0$ has essentially the same lifetime as that with a conductance of
$3 G_0$.  In this sense, the activation barrier appears to exhibit
{\em universal
mesoscopic fluctuations}:  in any conductance interval, there are very
short-lived wires (not shown in Table \ref{tab:lifetime}) with very
small activation barriers, while the longest-lived wires have activation
barriers of a universal size:
\begin{equation}
0\,< \,\Delta E_\infty\, \lesssim \, 0.7 
\left(\frac{\hbar^2 \sigma}{m_e}\right)^{1/2}\!\!\!\!\!,
\label{eq:Delta_E}
\end{equation}
where $\sigma$ is the surface tension and $m_e$ is the free-electron rest
mass.  The scaling with $\sigma$ in Eq.\ (\ref{eq:Delta_E}) follows
straightforwardly if one neglects the third term on the right hand side of
Eq.\ (\ref{eq:Vtilde}), which is a small correction that tends to destabilize
the wires, and is more important in gold than in sodium.

The lifetimes tabulated for sodium nanowires in Table \ref{tab:lifetime} 
exhibit a rapid decrease in the temperature interval between 75K and 100K.
This behavior can explain the observed temperature dependence of conductance
histograms for sodium nanowires,\cite{yanson99,yanson00,yanson01a}
which show clear peaks at conductances near the predicted values at 
temperatures up to 100K, but were not reported at higher temperatures.
A comparison of the lifetimes of sodium and gold nanowires listed in 
Table \ref{tab:lifetime} indicates that gold nanowires are much more stable,
as expected from the larger value of the surface tension
$\sigma_{\rm Au}= 5.9\, \sigma_{\rm Na}$.  This is consistent with the
observation that gold nanowires in particular, and noble metal nanowires
in general, are much more stable than alkali metal nanowires.
However, the calculated lifetimes of gold nanowires are not sufficient 
to account for the observed stability of gold nanowires at room temperature
and above.  This quantitative discrepancy may arise due to the neglect 
of d-electrons in our model
(except in as much as they enhance $\sigma$ compared to the free-electron 
value),
which are believed to play an important role in gold nanostructures.

\acknowledgments

J.B.\ and C.A.S.\ acknowledge support from NSF Grant No.\ DMR0312028.
D.L.S.\ acknowledges support from NSF Grant Nos.\ PHY0099484 and
PHY0351964.


\bibliography{spie-refs-final}   
\bibliographystyle{spiebib}   

\end{document}